
\def\aam #1 #2 #3 {{\sl Adv.\ Appl.\ Mech.} {\bf #1}, #2 (#3)}
\def\aap #1 #2 #3 {{\sl Adv.\ Appl.\ Prob.} {\bf #1}, #2 (#3)}
\def\ac #1 #2 #3 {{\sl Adv.\ Catal.} {\bf #1}, #2 (#3)}
\def\aces #1 #2 #3 {{\sl Adv.\ Chem.\ Eng.\ Sci.} {\bf #1}, #2 (#3)}
\def\acp #1 #2 #3 {{\sl Adv.\ Chem.\ Phys.} {\bf #1}, #2 (#3)}
\def\ae #1 #2 #3 {{\sl Ann.\ Eugenics} {\bf #1}, #2 (#3)}
\def\ah #1 #2 #3 {{\sl Adv.\ Hydrosci.} {\bf #1}, #2 (#3)}
\def\aichej #1 #2 #3 {{\sl AICHE J.} {\bf #1}, #2 (#3)}
\def\aj #1 #2 #3 {{\sl Astrophys.\ J.} {\bf #1}, #2 (#3)}
\def\ajp #1 #2 #3 {{\sl Am.\ J.\ Phys.} {\bf #1}, #2 (#3)}
\def\ams #1 #2 #3 {{\sl Ann.\ Math.\ Stat.} {\bf #1}, #2 (#3)}
\def\ap #1 #2 #3 {{\sl Adv.\ Phys.} {\bf #1}, #2 (#3)}
\def\apa #1 #2 #3 {{\sl Appl.\ Phys.\ A} {\bf #1}, #2 (#3)}
\def\apc #1 #2 #3 {{\sl Appl.\ Catal.} {\bf #1}, #2 (#3)}
\def\annp #1 #2 #3 {{\sl Ann.\ Phys.\ (N.Y.)} {\bf #1}, #2 (#3)}
\def\annpr #1 #2 #3 {{\sl Ann.\ Probab.} {\bf #1}, #2 (#3)}
\def\arfm #1 #2 #3 {{\sl Ann.\ Rev.\ Fluid Mech.} {\bf #1}, #2 (#3)}
\def\arpc #1 #2 #3 {{\sl Ann.\ Rev.\ Phys.\ Chem.} {\bf #1}, #2 (#3)}
\def\astech #1 #2 #3 {{\sl Aerosol Sci.\ Tech.} {\bf #1}, #2 (#3)}
\def\bcdg #1 #2 #3 {{\sl Ber.\ Chem.\ Dtsch.\ Ges.} {\bf #1}, #2 (#3)}
\def\bio #1 #2 #3 {{\sl Biometrika} {\bf #1}, #2 (#3)}
\def\bj #1 #2 #3 {{\sl Biophys.\ J.} {\bf #1}, #2 (#3)}
\def\ces #1 #2 #3 {{\sl Chem.\ Engr.\ Sci.} {\bf #1}, #2 (#3)}
\def\cf #1 #2 #3 {{\sl Combust.\ and Flame} {\bf #1}, #2 (#3)}
\def\cmp #1 #2 #3 {{\sl Commun.\ Math.\ Phys.} {\bf #1}, #2 (#3)}
\def\cp #1 #2 #3 {{\sl Chem.\ Phys.} {\bf #1}, #2 (#3)}
\def\cpam #1 #2 #3 {{\sl Commun.\ Pure Appl.\ Math.} {\bf #1}, #2 (#3)}
\def\crev #1 #2 #3 {{\sl Catal.\ Rev.} {\bf #1}, #2 (#3)}
\def\crII #1 #2 #3 {{\sl C. R. Acad.\ Sci.\ Ser.\ II} {\bf #1}, #2 (#3)}
\def\eul #1 #2 #3 {{\sl Europhys.\ Lett.} {\bf #1}, #2 (#3)}
\def\ic #1 #2 #3 {{\sl Icarus} {\bf #1}, #2 (#3)}
\def\iec #1 #2 #3 {{\sl Ind.\ Eng.\ Chem.} {\bf #1}, #2 (#3)}
\def\ijf #1 #2 #3 {{\sl Int.\ J.\ of Fracture} {\bf #1}, #2 (#3)}
\def\ijmpA #1 #2 #3 {{\sl Int.\ J.\ Modern Phys. A} {\bf #1}, #2 (#3)}
\def\ijmpB #1 #2 #3 {{\sl Int.\ J.\ Modern Phys. B} {\bf #1}, #2 (#3)}
\def\ijss #1 #2 #3 {{\sl Int.\ J.\ Solids Structures} {\bf #1}, #2 (#3)}
\def\jam #1 #2 #3 {{\sl J.\ Appl.\ Mech.} {\bf #1}, #2 (#3)}
\def\jap #1 #2 #3 {{\sl J.\ Appl.\ Phys.} {\bf #1}, #2 (#3)}
\def\japr #1 #2 #3 {{\sl J.\ Appl.\ Prob.} {\bf #1}, #2 (#3)}
\def\jaes #1 #2 #3 {{\sl J.\ Aerosol.\ Sci.} {\bf #1}, #2 (#3)}
\def\jas #1 #2 #3 {{\sl J.\ Atmos.\ Sci.} {\bf #1}, #2 (#3)}
\def\jasa #1 #2 #3 {{\sl J. Acous.\ Soc.\ Amer.} {\bf #1}, #2 (#3)}
\def\jc #1 #2 #3 {{\sl J.\ Catal.} {\bf #1}, #2 (#3)}
\def\jce #1 #2 #3 {{\sl J.\ Chem.\ Educ.} {\bf #1}, #2 (#3)}
\def\jcis #1 #2 #3 {{\sl J.\ Colloid Interface Sci.} {\bf #1}, #2 (#3)}
\def\jcg #1 #2 #3 {{\sl J.\ Crystal Growth} {\bf #1}, #2 (#3)}
\def\jcp #1 #2 #3 {{\sl J.\ Chem.\ Phys.} {\bf #1}, #2 (#3)}
\def\jcompp #1 #2 #3 {{\sl J.\ Comp.\ Phys.} {\bf #1}, #2 (#3)}
\def\jdep #1 #2 #3 {{\sl J.\ de Physique I} {\bf #1}, #2 (#3)}
\def\jdepl #1 #2 #3 {{\sl J. de Physique Lett.} {\bf #1}, #2 (#3)}
\def\jetp #1 #2 #3 {{\sl Sov.\ Phys.\ JETP} {\bf #1}, #2 (#3)}
\def\jetpl #1 #2 #3 {{\sl Sov. Phys.\ JETP Letters} {\bf #1}, #2 (#3)}
\def\jes #1 #2 #3 {{\sl J. Electrochem.\ Soc.} {\bf #1}, #2 (#3)}
\def\jfi #1 #2 #3 {{\sl J. Franklin Inst.} {\bf #1}, #2 (#3)}
\def\jfm #1 #2 #3 {{\sl J. Fluid Mech.} {\bf #1}, #2 (#3)}
\def\jgr #1 #2 #3 {{\sl J.\ Geophys.\ Res.} {\bf #1}, #2 (#3)}
\def\jif #1 #2 #3 {{\sl J. Inst.\ Fuel} {\bf #1}, #2 (#3)}
\def\jmo #1 #2 #3 {{\sl J. Mod. Opt.} {\bf #1}, #2 (#3)}
\def\jmp #1 #2 #3 {{\sl J. Math. Phys.} {\bf #1}, #2 (#3)}
\def\jms #1 #2 #3 {{\sl J. Memb.\ Sci.} {\bf #1}, #2 (#3)}
\def\josaA #1 #2 #3 {{\sl J. Opt.\ Soc.\ Am.\ A} {\bf #1}, #2 (#3)}
\def\josaB #1 #2 #3 {{\sl J. Opt.\ Soc.\ Am.\ B } {\bf #1}, #2 (#3)}
\def\jpa #1 #2 #3 {{\sl J. Phys.\ A} {\bf #1}, #2 (#3)}
\def\jpc #1 #2 #3 {{\sl J. Phys.\ C} {\bf #1}, #2 (#3)}
\def\jpd #1 #2 #3 {{\sl J. Phys.\ D} {\bf #1}, #2 (#3)}
\def\jpchem #1 #2 #3 {{\sl J. Phys.\ Chem.} {\bf #1}, #2 (#3)}
\def\jps #1 #2 #3 {{\sl J. Polymer.\ Sci.} {\bf #1}, #2 (#3)}
\def\jpsj #1 #2 #3 {{\sl J. Phys.\ Soc. Jpn.} {\bf #1}, #2 (#3)}
\def\jpso #1 #2 #3 {{\sl J. Power Sources} {\bf #1}, #2 (#3)}
\def\jsc #1 #2 #3 {{\sl J. Sci.\ Comp.} {\bf #1}, #2 (#3)}
\def\jsp #1 #2 #3 {{\sl J. Stat.\ Phys.} {\bf #1}, #2 (#3)}
\def\jtb #1 #2 #3 {{\sl J. Theor.\ Biol.} {\bf #1}, #2 (#3)}
\def\kc #1 #2 #3 {{\sl Kinet.\ Catal.\ (USSR)} {\bf #1}, #2 (#3)}
\def\macro #1 #2 #3 {{\sl Macromolecules} {\bf #1}, #2 (#3)}
\def\mclc #1 #2 #3 {{\sl Mol.\ Cryst.\ Liq.\ Cryst.} {\bf #1}, #2 (#3)}
\def\mubm #1 #2 #3 {{\sl Moscow Univ.\ Bull.\ Math.} {\bf #1}, #2 (#3)}
\def\nat #1 #2 #3 {{\sl Nature} {\bf #1}, #2 (#3)}
\def\npa #1 #2 #3 {{\sl Nucl.\ Phys.\ A} {\bf #1}, #2 (#3)}
\def\npb #1 #2 #3 {{\sl Nucl.\ Phys. B} {\bf #1}, #2 (#3)}
\def\PA #1 #2 #3 {{\sl PAGEOPH} {\bf #1}, #2 (#3)}
\def\pA #1 #2 #3 {{\sl Physica A} {\bf #1}, #2 (#3)}
\def\pBC #1 #2 #3 {{\sl Physica B \& C} {\bf #1}, #2 (#3)}
\def\pD #1 #2 #3 {{\sl Physica D} {\bf #1}, #2 (#3)}
\def\pams #1 #2 #3 {{\sl Proc.\ Am.\ Math.\ Soc.} {\bf #1}, #2 (#3)}
\def\pcps #1 #2 #3 {{\sl Proc.\ Camb.\ Philos.\ Soc.} {\bf #1}, #2 (#3)}
\def\pfA #1 #2 #3 {{\sl Phys.\ Fluids A} {\bf #1}, #2 (#3)}
\def\pla #1 #2 #3 {{\sl Phys.\ Lett. A} {\bf #1}, #2 (#3)}
\def\plb #1 #2 #3 {{\sl Phys.\ Lett. B} {\bf #1}, #2 (#3)}
\def\pmB #1 #2 #3 {{\sl Philos.\ Mag. B} {\bf #1}, #2 (#3)}
\def\pnas #1 #2 #3 {{\sl Proc.\ Natl.\ Acad.\ Sci.} {\bf #1}, #2 (#3)}
\def\pr #1 #2 #3 {{\sl Phys.\ Rev.} {\bf #1}, #2 (#3)}
\def\pra #1 #2 #3 {{\sl Phys.\ Rev.\ A} {\bf #1}, #2 (#3)}
\def\prb #1 #2 #3 {{\sl Phys.\ Rev.\ B} {\bf #1}, #2 (#3)}
\def\prc #1 #2 #3 {{\sl Phys.\ Rev.\C} {\bf #1}, #2 (#3)}
\def\prd #1 #2 #3 {{\sl Phys.\ Rev.\ D} {\bf #1}, #2 (#3)}
\def\pre #1 #2 #3 {{\sl Phys.\ Rev.\ E} {\bf #1}, #2 (#3)}
\def\prept #1 #2 #3 {{\sl Phys.\ Repts.} {\bf #1}, #2 (#3)}
\def\prk #1 #2 #3 {{\sl Prog.\ Reac.\ Kinetics} {\bf #1}, #2 (#3)}
\def\prl #1 #2 #3 {{\sl Phys.\ Rev.\ Lett.} {\bf #1}, #2 (#3)}
\def\prsl #1 #2 #3 {{\sl Proc.\ Roy.\ Soc.\ London Ser. A} {\bf #1}, #2 (#3)}
\def\pss #1 #2 #3 {{\sl Prog.\ Surf.\ Sci.} {\bf #1}, #2 (#3)}
\def\pt #1 #2 #3 {{\sl Powder Tech.} {\bf #1}, #2 (#3)}
\def\ptrs #1 #2 #3 {{\sl Phil.\ Trans.\ Roy.\ Soc., Ser. A} {\bf #1}, #2 (#3)}
\def\rjpc #1 #2 #3 {{\sl Russ.\ J. Phys.\ Chem.} {\bf #1}, #2 (#3)}
\def\rmp #1 #2 #3 {{\sl Rev.\ Mod.\ Phys.} {\bf #1}, #2 (#3)}
\def\rpp #1 #2 #3 {{\sl Rep.\ Prog.\ Phys.} {\bf #1}, #2 (#3)}
\def\sci #1 #2 #3 {{\sl Science} {\bf #1}, #2 (#3)}
\def\sciam #1 #2 #3 {{\sl Scientific American} {\bf #1}, #2 (#3)}
\def\SIAMjam #1 #2 #3 {{\sl SIAM J.\ Appl.\ Math.} {\bf #1}, #2 (#3)}
\def\spu #1 #2 #3 {{\sl Sov.\ Phys.\ Usp.} {\bf #1}, #2 (#3)}
\def\SPEre #1 #2 #3 {{\sl SPE Reservoir Eng.} {\bf #1}, #2 (#3)}
\def\ssc #1 #2 #3 {{\sl Sol.\ State.\ Commun.} {\bf #1}, #2 (#3)}
\def\ss #1 #2 #3 {{\sl Surf.\ Sci.} {\bf #1}, #2 (#3)}
\def\stec #1 #2 #3 {{\sl Sov.\ Theor.\ Exp.\ Chem.} {\bf #1}, #2 (#3)}
\def\tAIME #1 #2 #3 {{\sl Trans.\ AIME} {\bf #1}, #2 (#3)}
\def\tfs #1 #2 #3 {{\sl Trans.\ Faraday Soc.} {\bf #1}, #2 (#3)}
\def\tpa #1 #2 #3 {{\sl Theor.\ Prob.\ Appl.} {\bf #1}, #2 (#3)}
\def\tpm #1 #2 #3 {{\sl Transport in Porous Media} {\bf #1}, #2 (#3)}
\def\usp #1 #2 #3 {{\sl Sov.\ Phys. -- Usp.} {\bf #1}, #2 (#3)}
\def\wrr #1 #2 #3 {{\sl Water Resources Res.} {\bf #1}, #2 (#3)}
\def\zpb #1 #2 #3 {{\sl Z. Phys.\ B} {\bf #1}, #2 (#3)}
\def\zpc #1 #2 #3 {{\sl Z. Phys.\ Chem.} {\bf #1}, #2 (#3)}
\def\ztf #1 #2 #3 {{\sl Zh. Tekh.\ Fiz.} {\bf #1}, #2 (#3)}
\def\zw #1 #2 #3 {{\sl Z. Wahrsch.\ verw.\ Gebiete} {\bf #1}, #2 (#3)}

\def\gl{\mathrel{\raise1ex\hbox{$>$\kern-.75em\lower1ex\hbox{$<$}}}}
\def\lg{\mathrel{\raise1ex\hbox{$<$\kern-.75em\lower1ex\hbox{$>$}}}}
\def\gtwid{\mathrel{\raise.3ex\hbox{$>$\kern-.75em\lower1ex\hbox{$\sim$}}}}
\def\ltwid{\mathrel{\raise.3ex\hbox{$<$\kern-.75em\lower1ex\hbox{$\sim$}}}}
\def\sqr#1#2{{\vcenter{\hrule height.#2pt
      \hbox{\vrule width.#2pt height#1pt \kern#1pt
         \vrule width.#2pt}
      \hrule height.#2pt}}}

\overfullrule=0pt

\def\eg{\hbox{{\it e.\ g.}}}\def\ie{\hbox{{\it i.\ e.}}}\def\etc{{\it etc.}}
\def\vv{{\it vice versa}}\def\ea{\hbox{{\it et al.}}}

\def\ai{{\it ad infinitum}}


\def\leaderfill{\leaders\hbox to 1em{\hss.\hss}\hfill}

\def\CA{\hbox{{$\cal A$}}}
\def\CB{\hbox{{$\cal B$}}}

\def\CM{\hbox{{$\cal M$}}}

\def\CP{\hbox{{$\cal P$}}}

\def\ref#1{${}^{#1}$}
\newcount\eqnum \eqnum=0  
\newcount\eqnA\eqnA=0\newcount\eqnB\eqnB=0\newcount\eqnC\eqnC=0\newcount\eqnD\eqnD=0
\def\eqnoi{\global\advance\eqnum by 1\eqno(\the\eqnum)}
\def\eqnai{\global\advance\eqnum by 1\eqno(\the\eqnum{a})}
\def\eqnbi{\eqno(\the\eqnum{b})}
\def\eqnci{\eqno(\the\eqnum{c})}
\def\eqnoA{\global\advance\eqnA by 1\eqno(A\the\eqnA)}
\def\eqnoB{\global\advance\eqnB by 1\eqno(B\the\eqnB)}
\def\eqnoC{\global\advance\eqnC by 1\eqno(C\the\eqnC)}
\def\eqnoD{\global\advance\eqnD by 1\eqno(D\the\eqnD)}
\def\back#1{{\advance\eqnum by-#1 Eq.~(\the\eqnum)}}
\def\backa#1{{\advance\eqnum by-#1 Eq.~(\the\eqnum(a))}}
\def\backb#1{{\advance\eqnum by-#1 Eq.~(\the\eqnum(b))}}
\def\backc#1{{\advance\eqnum by-#1 Eq.~(\the\eqnum(c))}}
\def\backs#1{{\advance\eqnum by-#1 Eqs.~(\the\eqnum)}}
\def\backn#1{{\advance\eqnum by-#1 (\the\eqnum)}}
\def\backA#1{{\advance\eqnA by-#1 Eq.~(A\the\eqnA)}}
\def\backB#1{{\advance\eqnB by-#1 Eq.~(B\the\eqnB)}}
\def\backC#1{{\advance\eqnC by-#1 Eq.~(C\the\eqnC)}}
\def\backD#1{{\advance\eqnD by-#1 Eq.~(D\the\eqnD)}}
\def\last{{Eq.~(\the\eqnum)}}                     
\def\lasta{{Eq.~(\the\eqnum(a))}}                 
\def\lastb{{Eq.~(\the\eqnum(b))}}                 
\def\lastc{{Eq.~(\the\eqnum(c))}}                 
\def\lasts{{Eqs.~(\the\eqnum)}}                   
\def\lastn{{(\the\eqnum)}}                      
\def\lastA{{Eq.~(A\the\eqnA)}}\def\lastB{{Eq.~(B\the\eqnB)}}
\def\lastC{{Eq.~(C\the\eqnC)}}\def\lastD{{Eq.~(D\the\eqnD)}}
\newcount\refnum\refnum=0  
\def\refi{\smallskip\global\advance\refnum by 1\item{\the\refnum.}}

\newcount\rfignum\rfignum=0  
\def\rfigi{\medskip\global\advance\rfignum by 1\item{Figure \the\rfignum.}}

\newcount\fignum\fignum=0  
\def\figi{\global\advance\fignum by 1 Fig.~\the\fignum}

\newcount\rtabnum\rtabnum=0  
\def\rtabi{\medskip\global\advance\rtabnum by 1\item{Table \the\rtabnum.}}

\newcount\tabnum\tabnum=0  
\def\tabi{\global\advance\tabnum by 1 Table~\the\tabnum}

\newcount\secnum\secnum=0 
\def\chap#1{\global\advance\secnum by 1
\bigskip\centerline{\bf{\the\secnum}. #1}\smallskip\noindent}

\newcount\nlet\nlet=0
\def\numblet{\relax \global\advance\nlet by 1
\ifcase \nlet \ \or a\or b\or c\or d\or e\or
f\or g\or h\or i\or j\or k\or l\or m\or n\or o\or p
\or q \or r \or s \or t \or u \or v \or w \or x \or y
\or z \else .\nlet \fi}
\def\asubsec#1{\smallskip{\bf\centerline{(\numblet) #1}\smallskip}}
\def\tasubsec#1{{\bf\centerline{(\numblet) #1}\smallskip}}

\newcount\rslet\rslet=0
\def\romslet{\relax \global\advance\rslet by 1
\ifcase \rslet \ \or i\or ii\or iii\or iv\or v\or
vi\or vii\or viii\or ix\or x\or xi\or xii \else .\rslet \fi}

\newcount\rlet\rlet=0
\def\romlet{\relax \global\advance\rlet by 1
\ifcase \rlet \ \or I\or II\or III\or IV\or V\or
VI\or VII\or VIII\or IX\or X\or XI\or XII\or XIII\or XIV\or XV\or XVI
\or XVII \else .\rlet \fi}
\def\rchap#1{\bigskip\centerline{\bf{\romlet}. #1}\smallskip\noindent}

\def\pd#1#2{{\partial #1\over\partial #2}}      
\def\p2d#1#2{{\partial^2 #1\over\partial #2^2}} 
\def\t2d#1#2{{d^2 #1\over d #2^2}} 


\def\2kth{{$2k^{\rm th}$}}

\def\nth{{$n^{\rm th}$}}

\def\n-th{{$(n-1)^{\rm th}$}}

\def\N-th{{$(N-1)^{\rm th}$}}

\def\0th{$0^{\rm th}$}
\def\1st{$1^{\rm st}$}
\def\2nd{$2^{\rm nd}$}
\def\3rd{$3^{\rm rd}$}
\def\4th{$4^{\rm th}$}
\def\5th{$5^{\rm th}$}
\def\5th{$6^{\rm th}$}
\def\6th{$7^{\rm th}$}
\def\7th{$7^{\rm th}$}
\def\8th{$8^{\rm th}$}
\def\9th{$9^{\rm th}$}

\def\a{{\alpha}}
\def\b{{\beta}}
\def\G{\Gamma}

\def\l{\lambda}
\def\m{\mu}
\def\n{\nu}

\def\t{\tau}

\overfullrule=0pt
\magnification 1200
\baselineskip=15.1 true bp

\centerline{\bf Borderline Aggregation Kinetics in
``Dry'' and ``Wet'' Environments}
\bigskip
\centerline{\bf P.~L.~Krapivsky$\dag$ and S.~Redner}
\smallskip
\centerline{Center for Polymer Studies and Department of Physics}
\centerline{Boston University, Boston, MA 02215}
\vskip 0.6in

\centerline{ABSTRACT}

{\narrower\narrower\smallskip\noindent We investigate the kinetics of
constant-kernel aggregation which is augmented by either: (a)
evaporation of monomers from finite-mass clusters, or (b) continuous
cluster growth -- \ie, condensation.  The rate equations for these two
processes are analyzed using both exact and asymptotic methods.  In
aggregation-evaporation, if the evaporation is mass conserving, \ie, the
monomers which evaporate remain in the system and continue to be
reactive, the competition between evaporation and aggregation leads to
several asymptotic outcomes.  For weak evaporation, the kinetics is
similar to that of aggregation with no evaporation, while equilibrium is
quickly reached in the opposite case.  At a critical evaporation rate,
the cluster mass distribution decays as $k^{-5/2}$, where $k$ is the
mass, while the typical cluster mass grows with time as $t^{2/3}$.  In
aggregation-condensation, we consider the process with a growth rate for
clusters of mass $k$, $L_k$, which is: (i) independent of $k$, (ii)
proportional to $k$, and (iii) proportional to $k^\mu$, with $0<\mu<1$.
In the first case, the mass distribution attains a conventional scaling
form, but with the typical cluster mass growing as $t\ln t$.  When
$L_k\propto k$, the typical mass grows exponentially in time, while the
mass distribution again scales.  In the intermediate case of $L_k\propto
k^\mu$, scaling generally applies, with the typical mass growing as
$t^{1/(1-\mu)}$.  We also give an exact solution for the linear growth
model, $L_k\propto k$, in one dimension.

\bigskip
\bigskip\noindent
P. A. C. S. Numbers: 02.50.-r, 05.40.+j, 82.20.-w, 82.20.Mj

}

\vfill\eject

\rchap{INTRODUCTION}

Aggregation, fragmentation, and condensation underlie a variety of
non-equilibrium phenomena in nature [1-6].  In systems where only one of
these mechanisms is operative, the time-dependent cluster mass
distribution generally evolves to a scaling form in which the basic
variable is the ratio of the cluster mass to the typical mass.  These
scaling forms have been obtained by exact solutions, numerical
simulations, and by direct consistency checks of the scaling
description.  By these efforts a general understanding has been
developed for the connection between microscopic reaction details and
macroscopic features of the cluster distribution.

These approaches have also been successfully applied to processes where
the mechanisms of aggregation, fragmentation, and condensation are
simultaneously active.  One important example is aggregation in
combination with fragmentation, a process which arises naturally in
reversible polymerization [7].  Since the basic elements of aggregation
and fragmentation are manifestly opposed, their combined effect
generally leads to an equilibrium state in a closed system.
Characterizing the approach to and the detailed nature of this
equilibrium has been of basic interest [7-11].  Detailed balance
considerations can generally be applied to determine the nature of the
equilibrium state.  If $K_{ij}$ denotes the aggregation rate of
$c_i+c_j\to c_{i+j}$, where $c_k$ is the concentration of clusters of
mass $k$, and $F_{ij}$ denotes the fragmentation rate of $c_{i+j}\to
c_i+c_j$, then detailed balance gives
$K_{ij}\overline{c}_i\overline{c}_j=F_{ij}\overline{c}_{i+j}$ where
$\overline{c}_k$ is the steady-state concentration of $k$-mers.

On the other hand, the situation where the fragmentation matrix $F_{ij}$
has mostly zero elements naturally arises in polymer chain growth
kinetics [12,13].  For example, $k$-mers may be unstable to break-off of
monomers (``evaporation''), while all other fragmentation events are
forbidden.  By the nature of these restrictions, the process may be
viewed as aggregation in a ``dry'' environment.  For this system, the
aforementioned detailed balance argument can no longer determine the
equilibrium state.  One of our goals is to investigate the kinetics of
this composite aggregation-evaporation process in the rate equation
approximation.  We generally consider the situation where the
aggregation rate is independent of the masses of the two incident
clusters.  We further restrict ourselves to the interesting case of mass
conservation, where the monomers which have evaporated remain in the
system and continue to participate in the reaction.  If the evaporation
is sufficiently strong, its dominance over the effects of aggregation
results in an equilibrium whose properties are determined analytically.
In the opposite case where aggregation dominates, the typical cluster
mass increases indefinitely.  However, at a critical value of the
evaporation rate, the mass distribution decays as $k^{-5/2}$, while the
typical cluster mass grows with time as $t^{2/3}$.  These intriguing
features emerge from both exact solutions to the rate equations and
asymptotic arguments.

In a complementary direction, we also explore the kinetics of a combined
aggregation-condensation process in which each cluster grows at a
specified mass-dependent rate, in addition to the aggregation.  This
underlying growth can be viewed as arising from a uniform flux of
monomers which permeate the system.  In this sense, the composite
reaction can be viewed as aggregation in a ``wet'' environment.  Such a
process naturally arises in various contexts, such as the growth of
breath figures and thin film deposition and growth [6,14,15].  For
breath figures, in particular, theoretical models have generally
considered the growth rate of each droplet to be a specified function of
time.  This ultimately gives rise to much faster droplet growth in the
composite aggregation-condensation process.  In contrast, we consider a
droplet growth rate which depends only on the cluster size, and not
explicitly on the time.  We also employ the rate equations, a
description which may not necessarily apply to the stationary droplets
which are characteristic of breath figure systems.  Within our
approximation, we find a number of surprising features in the time
evolution of the cluster mass distribution for a mass-independent
aggregation rate.  When the condensation rate is independent of the
cluster mass, the distribution of cluster masses obeys conventional
scaling, but with the typical mass growing as $t\,\ln t$.  On the other
hand, for a condensation rate which is proportional to the mass, the
typical mass grows as $e^t$.  For a general mass-dependent condensation
rate which is proportional to $k^\mu$ (with $0<\mu<1$), the typical mass
grows as $t^{1/(1-\mu)}$.  For all three cases, that is, $0\leq\mu\leq
1$, scaling generally applies and this approach further predicts that
the mass distribution decays exponentially with mass.

A detailed treatment of these two cases of aggregation-evaporation and
aggregation-condensation is presented in the following two sections.

\rchap{KINETICS OF AGGREGATION-EVAPORATION}

The rate equations for  mass conserving aggregation-fragmentation are,
$$
\eqalign{
\dot c_k(t)=
{1\over 2}\mathop{{\sum}'}_{i,j}&K_{ij}\,c_i(t)c_j(t)-
c_k(t)\sum_{i=1}^\infty K_{ki}\,c_i(t)\cr
&+\left[L_{k+1}\, c_{k+1}(t)-L_k \,c_k(t)\right]
+\delta_{k,1}\sum_{i=1}^\infty L_i\, c_i(t).\cr}
\eqnai
$$
Here the overdot denotes the time derivative and $c_k(t)$ is the
concentration of clusters of mass $k$ at time $t$.  In this equation,
the first two terms account for the gain and loss of $k$-mers due to
aggregation, respectively.  The prime on the sum in the gain term
indicates the restriction of mass conservation, $i+j=k$.  In the
evaporation process, a $k$-mer produces a $(k-1)$-mer and a monomer at a
rate $L_k\equiv\lambda k^\mu$.  The gain and loss of $k$-mers because of
evaporation are described by the third and fourth terms of \lasta.
Finally, the last term accounts for monomer production as a result of
evaporation.  (If monomers were removed from the system by evaporation,
the last term would be absent.  This non-conservative process has
trivial kinetics in which the mass disappears exponentially in time.)

Let us now determine the conditions for which the system either reaches
equilibrium or evolves \ai.  We focus on the case where the aggregation
rates $K_{ij}$ are all equal (constant-kernel aggregation) and where the
evaporation rate is independent of the cluster mass ($\mu=0$).
Physically, this latter rate is appropriate for linear polymers with
evaporation possible only at the chain ends.  Although one can easily
generalize the discussion to mass-dependent aggregation and evaporation
rates, their relative influences are typically of different orders of
magnitude.  Consequently, it is relatively straightforward to anticipate
whether an equilibrium or a scaling distribution obtains.  However, for
mass-independent aggregation and evaporation, the competition between
these two influences is delicate and gives rise to surprisingly rich
kinetic behavior.

With the assumptions of constant reaction kernel and size-independent
evaporation, the rate equations simplify to
$$
\dot c_k(t)= {1\over 2}\mathop{{\sum}'}_{i,j}K\,c_i(t)c_j(t)-
c_k(t)\sum_{i=1}^\infty K\,c_i(t)+\lambda\left(c_{k+1}(t)-c_k(t)\right)
+\lambda\, \delta_{k,1}\sum_{i=1}^\infty c_i(t).
\eqnbi
$$
The evaporation rate $\lambda$ and reaction rate $K$ can be
absorbed by redefining the concentrations and time by
$c_k(t)\to 2\lambda c_k(t)$ and $t\to t/\lambda K$, leading to
$$
\dot c_k(t)=\mathop{{\sum}'}_{i,j}c_i(t)c_j(t)-2c_k(t)\sum_{i=1}^\infty c_i(t)
+\left(c_{k+1}(t)-c_k(t)\right)+\delta_{k,1}\sum_{i=1}^\infty c_i(t).
\eqnci
$$
For simplicity, consider a monomer-only initial condition,
$c_k(t=0)=\CM\delta_{k,1}$.  In this case, the total initial mass is the
only control parameter, with a large mass corresponding to a small
evaporation rate and \vv.

To gain insight into the kinetics, it is helpful to first write the
equations for moments of the mass distribution, $M_n(t)\equiv
\sum_{k\ge 1} k^n c_k(t)$.  By straightforward manipulations on \lastc,
these moments obey the equations,
$$
\eqalign{
\dot M_0(t)&= -M_0(t)^2+M_0(t)-c_1(t),\cr
\dot M_1(t)&=0,\cr
\dot M_2(t)&=2(M_1^2-M_1+M_0(t)),\cr
\dot M_3(t)&=3M_1+3(2M_1-1)M_2(t),\cr
\dot M_4(t)&=2M_0(t)-4M_1+6(M_2(t)+M_2(t)^2)+4(2M_1-1)M_3(t), \cr
\dot M_5(t)&=5M_1-10M_2(t)+10M_3(t)+20M_2(t)M_3(t)+5(2M_1-1)M_4(t), \cr
        &\qquad\qquad\qquad\qquad \vdots\cr}
\eqnoi
$$
For the monomer-only initial condition, $M_n(t=0)=\CM$ for all $n$.
{}From the equations for $M_n(t)$ for $n\ge 2$, it is clear that the
typical mass and higher moments grow indefinitely, if the initial mass
is sufficiently large.  In the complementary case, however, an
equilibrium state is possible.

More complete information about the kinetics can be obtained by
analyzing the rate equations themselves.  For this purpose,  we
introduce the generating function
$$
g(z,t)=\sum_{k=1}^\infty c_k(t)z^k.\eqnoi
$$
By multiplying the rate equation for each $c_k(t)$ by $z^k$ and summing
over all $k$, the generating function obeys
$$
\dot g(z,t)=g(z,t)^2-2g(z,t)M_0(t)+\left({g(z,t)\over z}-
c_1(t)\right)-g(z,t)+M_0(t)z.
\eqnai
$$
Here we use the equivalence $g(z=1,t)=M_0(t)$, with $M_0(t)$ the cluster
number density.  As is often the case in these types of systems,
it is more convenient to consider a modified generating function,
$h(z,t)\equiv g(z,t)-g(1,t)= g(z,t)-M_0(t)$, in which the value at
$z=1$ is subtracted off.  This generating function satisfies
$$
\dot h(z,t)= h^2(z,t)+{1-z\over z}~h(z,t)+{(1-z)^2\over z}M_0(t).\eqnbi
$$

While we have been unable to solve this differential equation in
general, the time independent solution is
$$
h(z)= {z-1\over 2z}\left(1-\sqrt{1-4M_0z}\right),\eqnoi
$$
with $h(z)\equiv h(z,t=\infty)$, $M_0\equiv M_0(t=\infty)$, and the sign
of the radical is fixed by requiring that $h(z)\to -M_0$ as $z\to 0$.
Once the value of $M_0$ is specified, the equilibrium solution, as well
as the conditions for equilibrium to exist can be determined.  From
\last, we conclude that equilibrium exists whenever $4M_0\leq 1$, while
if $4M_0>1$, the power series representation of the generating function
diverges and equilibrium is not reached.  The unknown quantity $M_0$ can
be related to the initial mass by the requirement that
$z\pd{h}{z}|_{z=1}$, which is the total mass of the system, equals the
initial mass $\CM$.  This leads to the condition
$\CM=\left(1-\sqrt{1-4M_0}\right)/2$, or equivalently, $M_0=\CM(1-\CM)$.
Since $M_0$ must be in the range $(0,1/4)$ for equilibrium to occur and
since $M_0$ must be an increasing function of $\CM$, the corresponding
constraint on the initial mass is $\CM\in (0,1/2)$.

The equilibrium properties of the cluster size distribution can be
obtained by expanding the generating function in \last\ in powers of $z$
for $4M_0\leq 1$.  Thus the equilibrium cluster mass
distribution, $c_k\equiv c_k(t=\infty)$, is given by
$$
c_k={1\over 4\sqrt{\pi}}\left[{\Gamma(k-{1\over 2})
\over{\Gamma(k+1)}}(4M_0)^k
-{\Gamma(k+{1\over 2})\over{\Gamma(k+2)}}(4M_0)^{k+1}\right],
\eqnoi
$$
where $\Gamma(z)$ is the gamma function.  When
$4M_0<1$ (equivalently, $\CM<1/2$), the
asymptotic behavior of $c_k$ is dominated by $(4M_0)^k$ and the mass
distribution decays exponentially in $k$.  On the other hand, when
$4M_0=1$ ($\CM=1/2$),  the mass
distribution has the power-law form
$$
c_k={3\over 8\sqrt{\pi}}~{\Gamma(k-{1\over 2})\over{\Gamma(k+2)}}
\propto k^{-5/2}.
\eqnoi
$$

The behavior of the moments of the mass distribution reflects the above
two possibilities.  Since all positive moments approach equilibrium
values for $\CM<1/2$, a recursive solution of the moment equations, Eq.~(2),
gives
$$
\eqalign{
M_0&=\CM(1-\CM),\cr
M_1&=\CM,\cr
M_2&={\CM\over{1-2\CM}},\cr
M_3&={3\CM(1-\CM)\over{2(1-2\CM)^3}}-{\CM(1+\CM)\over{2(1-2\CM)}},\cr
   &\qquad\qquad\qquad \vdots \cr}
\eqnoi
$$
On the other hand, for the limiting case of $\CM=1/2$ ($M_0=1/4$), the
power-law form of the cluster mass distribution leads to finite values
of the moments $M_n$ for $n<3/2$ and diverging values for $n\geq 3/2$.

When $\CM>1/2$, an equilibrium solution of the rate equations does not
exist and the transient behavior is of basic interest.  To understand
this behavior, it is helpful to first consider the equation of motion
for $M_2(t)$.  For $\dot M_2(t)$ to remain positive, $M_0(t)$ must
approach a constant whose value is greater than $\CM(1-\CM)$ as
$t\to\infty$ (Eq.~(2)).  This observation suggests that the cluster mass
distribution breaks up into two components for $\CM>1/2$.  In the
scaling component, the typical cluster mass $M_2(t)$ diverges as
$t\to\infty$.  Conversely, the remaining non-scaling (equilibrium)
component arises from the continued production of monomers by
evaporation which attempts to maintain the number of clusters $M_0$ and
the mass $\CM$ at their equilibrium values of 1/4 and 1/2, respectively.
Thus the scaling part of the mass distribution should contain the
remaining mass of $\CM-1/2$, while the number of clusters in this
sub-population should decay to zero.

Since the non-scaling component of the mass distribution is given by
\back1, the corresponding contribution to $M_n(t)$ is given by $\tilde
M_n(t)\sim \sum_{k\le t}k^{n-5/2}\sim t^{n-3/2}$ for $n>3/2$, while
$\tilde M_n(t)\to {\rm const.}<\infty$ for $n<3/2$.  These estimates are
based on the observation that the mass distribution approaches the
steady-state form $c_k\sim k^{-5/2}$ behind a leading edge which is
growing linearly in time, while there are essentially no clusters of
larger mass.  To determine the relative contribution of the scaling
component of the mass distribution to the moments, notice, from Eq.~(2),
that the typical cluster size $M_2(t)\propto t$.  This suggests that the
ratio $x\equiv k/t$ is the appropriate scaled mass with which one can
express the scaling component of the mass distribution as
$$
c_k(t)=t^{-2}\Phi(x).
\eqnoi
$$
The prefactor in \last\ is needed to have a finite mass in this scaling
part of the distribution.  If we are interested only in the asymptotic
behavior, then from \last, the contribution of this component of the
mass distribution to the \nth\ moment is
$$
M_n(t)=t^{n-1}A_n, \qquad {\rm where}\quad
A_n=\int_0^\infty dx\, x^n\,\Phi(x).
\eqnoi
$$

As a result of these estimates, it is evident that the moments with
$n>1$ are dominated by the scaling component of the mass distribution.
We therefore restrict ourselves to $n>1$ and insert the above scaling
expression into the moment equations.  Again, retaining only the
asymptotically relevant terms, Eqs.~(2) are transformed into a system of
equations for the amplitudes $A_n$ $$
(n-1)A_n=(2\CM-1)nA_{n-1}+\sum_{i=2}^{n-2}{n\choose i} A_i A_{n-i},
\quad n\ge 3.  \eqnoi
$$
To determine $A_2$, we use the third of Eqs.~(2) together
with the expected asymptotic value $M_0=1/4$.
This gives $A_2=2(\CM-{1\over 2})^2$.  To solve for the remaining $A_n$
we introduce the generating function
$$
\CA(z)=\sum_{j=2}^\infty {A_j\over j!}z^{j-1}.
\eqnoi
$$
It is then straightforward to transform the system
\backn1\ into the differential equation
$$
{d\CA\over dz}=\left(\CA+\CM-{1\over 2}\right)^2.
\eqnoi
$$
Solving \last\ yields
$$
\CA={\left(\CM-{1\over 2}\right)^2 z\over
1-\left(\CM-{1\over 2}\right)z}
\eqnoi
$$
and therefore
$$
M_n(t)=n!\left(\CM-{1\over 2}\right)^nt^{n-1}.
\eqnoi
$$
{}From the moments given by \lastn, it is evident that the
scaling component of the cluster mass distribution is
$$
c_k(t)={1\over \left(\CM-{1\over 2}\right)t^{2}}
\exp\left[-{k\over \left(\CM-{1\over 2}\right)t}\right].
\eqnoi
$$
We therefore arrive at the same expression for the mass distribution
as that which arises in pure aggregation with a constant
reaction rate [1-4].  The only difference is that the mass which
comprises the scaling component is equal to $\CM-{1\over 2}$.

To summarize, in the weak aggregation regime, $\CM<{1\over 2}$, the
mass distribution approaches the equilibrium form of
\back{10}\ at an exponential rate in time.  In the complementary strong
aggregation regime, $\CM>{1\over 2}$, the typical mass grows
linearly in time.  However, there is an anomalous enhancement in the
small-mass tail of the mass distribution which is of the form given
in \back9.  This residue arises from the continued re-introduction of
monomers into the system by evaporation.

At the critical point $\CM={1\over 2}$, a novel temporal behavior can be
anticipated in which the approach to equilibrium occurs at a power-law,
rather than an exponential rate.  Let us therefore hypothesize that the
number of clusters, $M_0(t)$, approaches its equilibrium value of 1/4 as
$t^{-\nu}$.  Employing this assumption in the moment equations gives the
series of relations $M_2(t)\sim t^{1-\nu}$, $M_3(t)={1+3t\over 2}\sim
t$, $M_4(t)\sim t^{3-2\nu}$, $M_5(t)\sim t^{3-\nu}$, \etc.  Since the
exponents of successive reduced moments should be equidistant within a
scaling description, the condition $M_4/M_3\sim M_3/M_2$, \eg, implies
$\nu={2\over 3}$.  This suggests the general formula,
$$
M_n(t)\sim t^{{2n\over 3}-1}
\eqnoi
$$
for $n>{3\over 2}$, while for $n<{3\over 2}$ the moments approach finite
values.

One can refine this analysis to obtain the amplitudes as well.  Writing
the expected asymptotic behavior as $M_n(t)=B_n\,t^{{2n\over 3}-1}$ for
$n>{3\over 2}$ and substituting into the moment equations, the $B_n$
satisfy recursion relations for integer $n$.   Defining $B_0$ via
$M_0(t)-{1\over 4}=B_0t^{-2/3}$, we then find $B_2=6B_0$, $B_3={3\over
2}$, and
$$
\Bigl({2n\over 3}-1\Bigr)B_n=\sum_{i=2}^{n-2}{n\choose i}B_iB_{n-i},
\eqnoi
$$
for $n\ge 4$.  While we could not solve \last\
generally, we can obtain partial information by a
generating function approach.  Indeed, introducing
$$
\CB(z)=\sum_{j=2}^\infty {B_j\over j!}z^{{2j\over 3}-1}
\eqnoi
$$
transforms \backs1\ into a Riccati equation
$$
{d\CB\over dz}=\CB^2+B_0z^{-2/3}+{1\over 4}.
\eqnoi
$$
While \last\ is insoluble when $B_2\ne 0$, several characteristics can
be established without an explicit solution.  In particular, the
structure of \last\ suggests that a solution exists for $z<z_c$ and it
has form of a simple pole, $\CB(z)\simeq (z_c-z)^{-1}$, in the vicinity
of $z=z_c$.  Inversion of this singularity yields the large-$n$ behavior
for every third amplitude, $B_{3n}\simeq 2(3n)!z_c^{-2n}$, while
$B_{3n+1}$ and $B_{3n+2}$ are asymptotically negligible compared to
$B_{3n}$.  The location of the singularity, $z_c$, depends on the
unknown amplitude $B_0$ and should be established from a more
comprehensive treatment which, in turn, is equivalent to finding a
complete solution of \last.

\rchap{KINETICS OF AGGREGATION-CONDENSATION}
\tasubsec{Mass Independent Growth Rate ($L_k={\rm const.}$)}

We now consider the complementary reaction where bimolecular aggregation
(with mass-independent aggregation rates), is supplemented by
single-cluster growth in which a cluster of mass $k$ grows at a rate
$L_k\propto k^{\mu}$.  We wish to understand how this additional growth
influences the kinetics of the underlying aggregation within the rate
equations.  Let us first investigate this composite
aggregation-condensation process for a size-independent growth rate,
$L_k=\lambda$.  The rate equations become
$$
\dot c_k(t)=\mathop{{\sum}'}_{i,j}c_i(t)c_j(t)-2c_k(t)\sum_{i=1}^\infty
c_i(t) +\lambda\left(c_{k-1}(t)-c_k(t)\right).  \eqnoi
$$
To gain qualitative insight into the asymptotic behavior, we begin by
solving for the first few moments of the mass distribution.  We then
present a complete solution for the mass distribution, from which the
asymptotic behavior may be extracted.

{}From the above rate equations, the moments evolve according to
$$
\eqalign{
\dot M_0(t)&= -M_0(t)^2,\cr
\dot M_1(t)&=\lambda M_0(t),\cr
\dot M_2(t)&=2M_1(t)^2+\lambda(2M_1(t)+M_0(t)),\cr
   & \qquad\qquad\vdots \cr}
\eqnoi
$$
subject to the initial condition $M_n(t=0)=1$ for all $n$.  We also set the
initial density equal to unity so the condensation rate $\lambda$ is the
only control parameter.  Solving for the moments
successively yields
$$
\eqalign{
M_0(t)=&~{1\over 1+t},\cr
M_1(t)=&~1+\lambda\ln(1+t),\cr
M_2(t)=&~1+\lambda\ln(1+t)+(4\lambda-2\lambda^2)(1+t)\ln(1+t) \cr
        &~~~~~~~~~~~~ +2\lambda^2(1+t)\ln^2(1+t)+2(1-\lambda+\lambda^2)t, \cr
    &\qquad\qquad\qquad\qquad\vdots \cr}
\eqnoi
$$
Although the exact expressions for $M_n(t)$ become
cumbersome as the index $n$ grows, the asymptotic behavior is simply
$$
M_n(t)\sim n!\,\lambda^n\, t^{n-1}\,(\ln t)^n.
\eqnoi
$$

To solve the full rate equations we again introduce the generating
function $g(z,t)=\sum_{k=1}^\infty c_k(t)z^k$, which reduces an infinite
set of rate equations \backn3\ to the differential equation
$$
\dot g(z,t)=g(z,t)^2-{2\over 1+t}~g(z,t)+\l(z-1)g(z,t).
\eqnoi
$$
Notice that $g(z,t)^{-1}$ satisfies a linear inhomogeneous differential
equation whose solution is
$$
g(z,t)={ze^{\l(z-1)t}\over{(1+t)^{2}}}\left[1-z\int_0^t
{d\tau\over (1+\tau)^2}\,e^{\l(z-1)\tau}\right]^{-1}.
\eqnoi
$$
We now determine the asymptotic behavior associated with this generating
function.  This information will also suggest the asymptotic form for
the mass distribution for general rates of aggregation and condensation,
a system for which an analytical solution cannot be found.  First, notice
that the densities $c_k(t)$ which make a non-zero contribution to the
generating function are those whose mass is in the range $0<(1-z)k\ltwid
1$.  If we further assume that $1-z\gg (\l t)^{-1}$, then the integral
in \last\ can be approximated by replacing the exponential by unity;
this is asymptotically correct over the domain of integration. The
generating function now becomes
$$
g(z,t)\cong(1+t)^{-2}e^{-\l t}\,{z\,e^{\l tz}\over 1-z}.
\eqnoi
$$
Expanding \last\ gives,
$$
c_{k+1}(t)\cong {(\l t)^k\over k!}t^{-2}e^{-\l t},
\qquad {\rm for} \quad 0\leq k\ll \l t.
\eqnoi
$$
Thus for the range $k\ll \l t$, the mass distribution is
Poissonian; however, the distribution cannot be written in the
conventional scaling form $t^{-\a}\Phi(k/t^\b)$.

On the other hand, for sufficiently large $k$, the mass distribution
does exhibit scaling.  To determine an appropriate mass scale we expand
the exponent $e^{\l(z-1)\tau}$, compute the integral in the right-hand
side of \back2, and then asymptotically balance the various terms. The
nontrivial scaling limit arises when $(1-z)\sim (t\ln t)^{-1}$ and
suggests that the appropriate scaling variable is
$$
\zeta=(1-z)\l t\ln t,
\eqnoi
$$
instead of the original variables $z$ and $t$.

In the scaling limit, $t\to \infty$ and $1-z\to 0$ with $\zeta$ kept
fixed, the generating function simplifies to
$$
g(z,t)\simeq t^{-1}(1+\zeta)^{-1},
\eqnoi
$$
and the mass distribution approaches the scaling form
$$
c_k(t)\simeq \phi(t)\Phi(x),
\qquad {\rm with}\quad x={k\over \l t\ln t},
\eqnoi
$$
with the prefactor $\phi(t)$ and scaling function $\Phi(x)$ to be
determined.  Making use of \backs2\ and \lastn, we express the
generating function in terms of $\phi(t)$ and $\Phi(x)$ as,
$$
\eqalign{
g(z,t)=&\sum z^kc_k(t)=\sum \left(1-{\zeta\over \l t\ln t}\right)^kc_k(t)\cr
       &\simeq ~\l t\ln t~\phi(t) \int_0^\infty \Phi(x) e^{-x\zeta} dx.\cr}
\eqnoi
$$
Finally, by comparing \backs2\ and \lastn\ and performing the inverse
Laplace transform, the prefactor $\phi(t)$ and the scaling
function $\Phi(x)$ are
$$
\phi(t)={1\over \l t^2\ln t},
\qquad {\rm and}\quad \Phi(x)=e^{-x}.
\eqnoi
$$
The scaling solution of \backs2\ and \lastn\ agrees with asymptotic
expression for the moments \backn9.

For completeness, we also investigate the large-mass tail of the mass
distribution, $k\gg \l t\ln t$.  The analysis is similar to that given
above so we merely cite the result
$$
c_k(t)\sim \left(1-{1\over \l t\ln t}\right)^k.
\eqnoi
$$
Thus the mass distribution function does not scale in both the small-
and large-mass tails.  Formally, the condensation process governs the
small-mass tail of the distribution, as well as the overall mass.
Conversely, the form of the distribution in the scaling region and in
the large-mass tail is determined solely by the aggregation process.

\asubsec{Growth Rate Proportional to the Mass ($L_k\propto k$)}

We now consider the case of a condensation rate which is linear in the
mass, \ie, the rate of $c_k\to c_{k+1}$ equals $\l k$.
The corresponding rate equations are
$$
\dot c_k(t)=\mathop{{\sum}'}_{i,j}c_i(t)c_j(t)-2c_k(t)\sum_{i=1}^\infty c_i(t)
+\lambda\left[(k-1)c_{k-1(t)}-kc_k(t)\right].
\eqnoi
$$
Employing the generating function $g(z,t)$, \last\ becomes
$$
\pd{}{t}g(z,t)=g(z,t)^2-2g(z,t)M_0(t)+\l z(z-1)\pd{}{z}g(z,t).
\eqnoi
$$
Notice that the number of clusters $M_0(t)\equiv g(z=1,t)$ still satisfies
$\dot M_0(t)=-M_0(t)^2$;
hence, $M_0(t)=(1+t)^{-1}$.  Introducing again the modified generating
function, $h(z,t)=g(z,t)-M_0(t)$, transforms \last\ into a linear
equation for $h(z,t)^{-1}$
$$
\pd{}{t}h(z,t)^{-1}+\l z(1-z)\pd{}{z}h(z,t)^{-1}+1=0.
\eqnoi
$$

By introducing $w={1\over\lambda}\ln{z\over{1-z}}$, \last\ becomes a
first-order wave equation in the variables ($w,t$).  This equation
further simplifies by transforming from $(w,t)$ to $ u=t+w$ and $v=t-w$
to yield
$$
\pd{}{u}h(u,v)^{-1}=-{1\over 2},
\eqnoi
$$
with solution $h(u,v)^{-1}=-{u\over 2}+f(v)$.  Here $f(v)$ is fixed by
the initial conditions. For example, for monodisperse monomer-only initial
conditions, we obtain
$$
h(u,v)^{-1}=-1-{u+v\over 2}-e^{-\l v}=-1-t-e^{-\l t}{z\over 1-z}.
\eqnoi
$$
By now expanding $g(z,t)=h(z,t)+{1\over 1+t}$ in powers of $z$
the exact concentrations are,
$$
c_k(t)={e^{-\l t}\over (1+t)^2}\left(1-{e^{-\l t}\over 1+t}\right)^{k-1}.
\eqnoi
$$
In the scaling region, $k\to \infty$ and $t\to \infty$ with
$x=k/te^{\l t}$ finite, \last\ has the asymptotic form
$$
c_k(t)\simeq t^{-2}e^{-\l t}e^{-x}.
\eqnoi
$$

Notice that the scaling solution for the mass distribution in both the
cases of $\m=0$ and $\m=1$ may be written as
$$
c_k(t)\simeq {M_0(t)^{2}\over M_1(t)}e^{-x}, \qquad{\rm with}
\quad x={M_0(t)\over M_1(t)}k.
\eqnoi
$$
Although the mass density, $M_1(t)$, has a very different time
dependence for these two cases, the respective scaling functions are the
same and in fact identical to that in pure aggregation with a constant
aggregation rate [1-4].

\asubsec{General Mass Dependent Growth Rate ($L_k\propto k^\mu$)}

We now turn to aggregation-condensation for a general homogeneous
mass-dependent cluster growth rate, $L_k=\lambda k^\mu$, with $0<\m<1$.
Although it does not appear possible to solve the governing rate
equations, one expects the scaling form of \last\ to hold in the scaling
region.  This assumption reduces the problem to finding the first two
moments, $M_0(t)$ and $M_1(t)$.  The former task is trivial since the
condensation process does not alter the evolution of the number of
clusters, and $M_0(t)=(1+t)^{-1}$.  On the other hand, $M_1(t)$ is
determined by $\dot M_1(t)=\l M_\m(t)$ which is coupled to an undetermined
moment.  Fortunately, in the long-time limit we can use the scaling form
\lastn\ to estimate $M_\m(t)$ as
$$
M_\m(t)=\sum_{k=1}^\infty k^\m c_k(t) \simeq
\left({M_1(t)\over M_0(t)}\right)^{\m+1}{M_0(t)^2\over M_1(t)}
\int_0^\infty dx\, x^\m e^{-x}
=\G(1+\m)M_1(t)^\m M_0(t)^{1-\m}.
\eqnoi
$$
Thus asymptotically $\dot M_1(t)\simeq \l \G(1+\m) t^{\m-1}M_1(t)^\m$,
which may be solved to yield
$$
M_1(t)\simeq At^{\m\over 1-\m}, \qquad{\rm with}\quad
A=\left[\l(1-\m)\G(\m)\right]^{1\over 1-\m}.
\eqnoi
$$

It is instructive to compare this asymptotic result for the
typical cluster size,
$$
S(t)={M_1(t)\over M_0(t)}=At^{1\over 1-\m},
\eqnoi
$$
with a naive estimate provided by considering growing, but {\it
noninteracting}, \ie, non-aggregating, clusters.  This latter estimate
follows from $\dot S(t)=\l S(t)^\m$, which implies
$$
S(t)=A_0t^{1\over 1-\m}, \qquad {\rm with}\quad
A_0=\left[\l(1-\m)\right]^{1\over 1-\m}.
\eqnoi
$$
Therefore the system with growing but non-interacting droplets provides
the correct exponent of the time dependence for the droplet growth rate
in the interacting system.  However, the corresponding prefactor $A_0$
is slightly smaller than that of the interacting system.

\asubsec{Scaling Approach for Low Spatial Dimension}

For diffusion-controlled aggregation, the above mean-field approaches
are typically not applicable for spatial dimension $d\leq 2$ (see, \eg,
[16,17]).  However, for the aggregation-condensation process with a
homogeneous growth rate, $L_k\propto k^\m$ ($0\leq \m<1$), it is
possible to infer partial results for $d\leq2$ by applying scaling and
exploiting known results.  In particular, for diffusion-controlled
aggregation with a mass-independent
cluster diffusivity, the density of clusters (which is not altered by
the condensation process) is [16,17]
$$
M_0(t)\sim \cases {t^{-d/2}, & $d<2$;\cr
               {\ln t\over t}, & $d=2$.\cr}
\eqnoi
$$
We now again assume that asymptotically the mass distribution
approaches the scaling form
$$
c_k(t)\simeq {M_0(t)^2\over M_1(t)}\Phi_d(x), \qquad{\rm with}\quad
x={M_0(t)\over M_1(t)}k,
\eqnoi
$$
with a general $d$-dependent scaling function $\Phi_d(x)$.  The mass
density, $M_1(t)$, is determined from $\dot M_1(t)=\l M_\m(t)$, where the
moment $M_\m(t)$ is estimated to be (following the steps of the previous
subsection),
$$
M_\m(t)\simeq M_0(t)^{1-\m}M_1(t)^\m \int_0^\infty dx\, x^\m \,\Phi_d(x).
\eqnoi
$$
Ignoring numerical factors we solve for the mass density to obtain
$$
M_1(t)\sim \cases {t^{{1\over 1-\m}-{d\over 2}}, & $d<2, ~0\leq\m<1$;\cr
               t^{\m\over 1-\m}\ln t, & $d=2, ~0<\m<1$;\cr
               \ln^2 t, & $d=2, ~\m=0$.\cr}
\eqnoi
$$
Finally, combining \backs3, \backn2, and \lastn\ yields
$$
c_k(t)\sim \cases {t^{-{1\over 1-\m}-{d\over 2}}~\Phi_d(x),
                                         & $d<2, ~0\leq\m<1$;\cr
               t^{-{2-\m\over 1-\m}}\ln t~\Phi_2(x), & $d=2, ~0<\m<1$;\cr
               t^{-2}~\Phi_2(x), & $d=2, ~\m=0$.\cr}
\eqnoi
$$
in all three cases, the scaling variable is $x=kt^{-1/(1-\m)}$.

\asubsec{Exact Solution in One Dimension for $L_k\propto k$}

We now solve the aggregation-condensation process in one dimension when
the cluster growth rate is linear in the mass ($L_k\propto k$) and the
diffusivity of each cluster is mass independent, a case particularly
amenable to exact analysis.  The model is defined on the real line which
is populated by point clusters which: (i) diffuse with a mass
independent diffusivity $D$, (ii) grow with a rate proportional to their
masses, and (iii) aggregate irreversibly whenever two clusters meet.  A
convenient way to treat this problem analytically is to introduce [18]
the quantity $P_k(x,t)$, which is the probability that the total mass of
all clusters contained in an interval of length $x$ at time $t$ is equal
to $k$.  The evolution of the $P_k(x,t)$ are uncoupled and each such
function obeys the diffusion equation [18].  If we account for
condensation, the $P_k(x,t)$ now evolve according to
$$
\pd{}{t}P_k(x,t)=2D\p2d{}{x}P_k(x,t)+\l\left[(k-1)P_{k-1}-kP_k\right].
\eqnoi
$$
The second term clearly accounts for the change in mass in the interval
due to the cluster growth process.  The cluster concentrations,
$c_k(t)$, are then found from
$$
c_k(t)={\partial \over \partial x}P_k(x=0,t).
\eqnoi
$$
Note also the sum rule, $\sum_{k=0}^\infty P_k(x,t)\equiv 1$,
which allows us to focus on the $P_k$'s with $k\geq 1$.

To solve this system, we introduce the generating function,
$$
\CP(z,x,t)=\sum_{k=1}^\infty z^k P_k(x,t),
\eqnoi
$$
to transform \back2\ into
$$
{\partial \CP\over \partial t}=2D{\partial^2 \CP\over \partial x^2}
+\l z(z-1){\partial \CP\over \partial z}.
\eqnoi
$$
Note that the obvious boundary condition, $P_k(x=0)=\delta_{k0}$,
translates to the condition on the generating function, $\CP(x=0)=0$.
If initially the system is composed of monomers which have a Poissonian
distribution of positions with density unity, the corresponding
initial condition is
$$
P_k(t=0)={x^k\over k!}e^{-x}, \qquad \CP(t=0)=e^{-x(1-z)}-e^{-x}.
\eqnai
$$
Although the variable $x$ is defined only on the half-line $x\geq 0$, it
proves useful to consider \back1\ for $-\infty<x<\infty$,
and impose the initial condition
$$
\CP(t=0)=e^{x}-e^{x(1-z)}, \qquad {\rm for}\quad x<0.
\eqnbi
$$
With this antisymmetric initial data, the boundary condition $\CP(x=0)=0$
is manifestly satisfied.

Employing the auxiliary variables $u=t+{1\over\lambda}\ln{z\over{1-z}}$
and $v=t-{1\over\lambda}\ln{z\over{1-z}}$ introduced earlier, we further
simplify \back1\ to the diffusion equation
$$
{\partial \CP\over \partial u}=D{\partial^2 \CP\over \partial x^2},
\eqnoi
$$
subject to the initial conditions
$$
\eqalign{
\CP\big|_{u=-v}&=e^{-x/(1+e^{-\l v})}-e^{-x},
\qquad {\rm for}\quad x>0, \cr
\CP\big|_{u=-v}&=e^{x}-e^{x/(1+e^{-\l v})},
\qquad {\rm for}\quad x<0. \cr}
\eqnoi
$$
Solving \back1\ with the above initial conditions yields
$$
\eqalign{
\CP(u,v,x)={1\over \sqrt{4\pi D(u+v)}}
&\int_0^\infty d\xi e^{-\xi}
\left[\exp\left({\xi\over 1+e^{\l v}}\right)-1\right] \cr
&\times \left[\exp\left(-~{(x-\xi)^2\over 4D(u+v)}\right)-
\exp\left(-~{(x+\xi)^2\over 4D(u+v)}\right)\right].\cr}
\eqnoi
$$
Returning to the original variables we find
$$
\eqalign{
\CP(z,x,t)={1\over \sqrt{8\pi Dt}}
&\int_0^\infty d\xi e^{-\xi}
\left[\exp\left({\xi\over 1+e^{\l t}{1-z\over z}}\right)-1\right]\cr
&\times \left[\exp\left(-~{(x-\xi)^2\over 8Dt}\right)-
\exp\left(-~{(x+\xi)^2\over 8Dt}\right)\right].\cr}
\eqnoi
$$
Combining \last\ with \back7\ gives
$$
\sum_{k=1}^\infty z^kc_k(t)={1\over \sqrt{32\pi (Dt)^3}}
\int_0^\infty d\xi~\xi \exp\left(-\xi-{\xi^2\over 8Dt}\right)
\left[\exp\left({\xi\over 1+e^{\l t}({1-z\over z})}\right)-1\right].
\eqnoi
$$
\last\ is the formal solution to the problem -- for example, the
concentrations can be found by expanding the right-hand side
in powers of $z$.  However, explicit expressions for the concentrations
$c_k(t)$ are cumbersome, even for small $k$.  In contrast, the
moments of the mass distribution have simpler form.  For example,
$$
M_0(t)={1\over \sqrt{32\pi (Dt)^3}}
\int_0^\infty d\xi~\xi \exp\left(-{\xi^2\over 8Dt}\right)
\left(1-e^{-\xi}\right)\simeq {1\over \sqrt{2\pi Dt}} ,
\eqnoi
$$
in agreement with the known result [19].  Furthermore, the mass density
is given by $M_1(t)=e^{\l t}$.  This is expected since $M_1(t)$
satisfies $\dot M_1(t)=\l M_1(t)$.  These two moments provide a useful
consistency check of our exact results.

\rchap{SUMMARY AND DISCUSSION}

Our primary result was to elucidate the broad range of phenomenology
which arises from the combined effects of: (a) aggregation with
evaporation, and (b) aggregation with condensation.  In the former case,
the interesting situation is that of mass independent rates of
aggregation and evaporation in a mass conserving system.  An equilibrium
state is reached for sufficiently strong evaporation, while the kinetics
is essentially identical to that of pure aggregation when the
evaporation is relatively weak.  At a critical evaporation rate, there
is power-law kinetics in which the typical cluster size grows as
$t^{2/3}$, while the mass distribution decays with mass $k$ as
$k^{-5/2}$.  An essential ingredient in these results is that the
aggregation and evaporation rates are of the same order, so that their
competition leads to interesting manifestations.  Related transition
behavior was obtained for combined aggregation-evaporation by Virgil
\ea\ [10], but with both a mass-dependent aggregation rate (proportional
to the product of the cluster masses) and evaporation rate (proportional
to the cluster mass).  For this latter system, the effects of
evaporation and aggregation are, in some sense,  of the same order
of magnitude, leading to a transition between equilibrium and gelation
for a critical value of the ratio of the two rates and also distinct
power-law behavior at the transition.  Another not entirely unrelated
example where microscopic effects influence the approach to equilibrium
occurs in the one-dimensional reversible reaction $A+A\leftrightarrow A$
[20].  In this case, the relaxation has different functional forms
depending on the ratio of the initial density to the final equilibrium
density.

For aggregation with condensation, the cluster growth enhances the
aggregation, as expected.  When the growth rate is independent of
cluster mass, this enhancement is relatively weak, however, and the
typical cluster mass grows with time as $t\ln t$, compared to a linear
growth for aggregation with no condensation.  Conversely when the growth
rate is proportional to the mass, the typical mass grows exponentially
in time.  In the intermediate case of a growth rate for clusters of mass
$k$ given by $L_k\propto k^\mu$, with $0<\mu<1$, a scaling approach
indicates that the typical cluster mass grows as $t^{1/(1-\mu)}$, while
the scaling function in the mass distribution is a pure exponential for
$0\leq\mu\leq 1$.

Our aggregation-condensation model is related to breath figures [14,15]
as well as to other models of droplet growth and coalescence [6].  In
these latter systems, the growth rate of individual droplets is
generally a specified function of time, rather than of the mass.  If one
assumes that the mass of individual droplets grow $t^{\alpha}$, then the
typical cluster mass in the combined aggregation-condensation system
grows at a faster rate of $t^{\beta}$, with $\b=D\alpha/(D-d)$.  Here
$D$ is the spatial dimension of the droplets and $d$ (which must be
smaller than $D$) is the dimensionality of the substrate.  It would
interesting to understand what relation, if any, exists between these
results and our model of aggregation with explicit mass-dependent
cluster growth.

The rate equation approach is expected to provide the correct asymptotic
behavior for aggregation-condensation when the spatial dimension $d$ is
greater than 2 [16,17].  To extend our understanding to $d\leq 2$, a
simple-minded approach was developed which suggests that conventional
scaling continues to apply.  Furthermore, we derived an exact solution
for aggregation-condensation in one dimension with a growth rate linear
in the cluster mass, $L_k\propto k$, and with a mass-independent
diffusivity.  The scaling predictions are in satisfying agreement with
this exact solution.
\vfill\eject
\rchap{ACKNOWLEDGMENTS}

We thank R. M.~Ziff for helpful correspondence and suggestions.  We also
gratefully acknowledge ARO grant \#DAAH04-93-G-0021 and NSF grant
\#DMR-9219845 for partial support of this research.


\rchap{REFERENCES}

\item{$\dag$} Present address:  Courant Institute of Mathematical
Sciences, New York University, New York, NY~ 10012

\refi R. L. Drake in {\sl Topics in Current Aerosol Research}, eds.\
      G. M. Hidy and J. R. Brock (Pergamon, New York, 1972), Vol.\ 3 Pt.\ 2.

\refi S.~K.~Friedlander, {\sl Smoke, Dust and Haze:
      Fundamentals of Aerosol Behavior} (Wiley, New York, 1977).

\refi R. M. Ziff, in {\sl Kinetics of Aggregation and Gelation}, eds.\
      F. Family and D. P. Landau (North-Holland, Amsterdam, 1984).

\refi M.~H.~Ernst, in {\sl Fundamental Problems in Statistical
      Physics VI}, ed. E.~G.~D.~Cohen (Elsevier, New York, 1985).

\refi S.~Redner, in {\sl Statistical Models for the Fracture of
      Disordered Media}, eds. H.~J.~Herrmann and S.~Roux
      (North-Holland, 1990).

\refi P. Meakin, \rpp 55 157 1992 .

\refi P.~J.~Blatz and A.~V.~Tobolsky, \jpchem 49 77 1945 .

\refi P.~G.~J.~van Dongen and M.~H.~Ernst, \jpa 16 L327 1983 .

\refi M.~Aizenman and T.~A.~Bak, \cmp 65 203 1979 .

\refi F.~Family, P.~Meakin, and J.~M.~Deutch, \prl 57 727 1986 .

\refi C.~M.~Sorensen, H.~X.~Zhang, and T.~W.~Taylor, \prl 59 363 1987 .

\refi T.~E.~Ramabhadran, T.~W.~Peterson, and J.~H.~Seinfeld,
      \aichej 22 840 1976 .

\refi R.~D.~Vigil, R.~M.~Ziff, and B.~Lu, \prb 38 942 1988 .

\refi D. Beysens and C. M. Knobler, \prl 57 1433 1986 .

\refi T. M. Rogers, K. R. Elder, and R. C. Desai, \pra 38 5305 1988 .

\refi K.~Kang and S.~Redner, \pra 30 2833 1984 .

\refi L. Peliti, \jpa 19 L365 1986 .

\refi B.~R.~Thompson, \jpa 22 879 1989 .

\refi J.~L.~Spouge, \prl 60 871 1988 .

\refi D. ben-Avraham, M. A. Burschka, and C. R. Doering, \jsp 60 695
1990 .


\vfill\eject\bye